# Genomic Bottleneck, Neural Scaling, and the Emergence of Intelligence: A Theoretical Perspective on Biological and Artificial Neural Systems

*Dr. Arifa Khan[1] and Dr. Suki Venkatesan[2]**

[1] AI/ML Computational science Manager, Chennai, Indiasss

[2] CQRL Bits LLP, Vignesh Avenue, Selaiyur, Chennai, India

*Corresponding author: Dr. Arifa Khan*

Email: ak7641@srmist.edu.in

Dr. Suki Venkatesan: suki@cqrl.in

## Abstract

This study explores the relationship between biological neural evolution and the development of modern computational architectures through the lens of the genomic bottleneck hypothesis. We examine how the low-latency neural systems observed in insects, birds, reptiles, and mammals may be conceptually compared with the evolution of low-latency central processing units (CPUs) in computer architecture. The increasing demand for high-throughput visual processing contributed to the development of graphics processing units (GPUs), which in turn enabled large-scale parallel computation and the emergence of deep learning systems. Drawing upon insights from evolutionary biology, neuroscience, and artificial intelligence, the paper proposes that the expansion of the human brain may be interpreted as a response to information-processing constraints imposed by genomic bottlenecks, resulting in highly trainable and adaptable neural architectures. The development of language, writing, and external knowledge repositories further extended human cognitive capabilities beyond biological memory limitations, in a manner consistent with the extended mind thesis [1]. The study argues that contemporary digital ecosystems, including social media, large-scale online corpora, and GPU-accelerated transformer models, represent a continuation of this long-term process of cognitive extension. Furthermore, similarities between human collective intelligence and large language models are discussed with respect to bias, memorization, hallucination, and information propagation. The paper contributes an interdisciplinary theoretical framework linking neural scaling, computational evolution, and the emergence of generative artificial intelligence, while highlighting potential directions for mitigating undesirable behaviors in both human and artificial information systems.

## 1. Introduction

In an earlier study [2], we explored how the collective evolution of writing, printing, libraries, and digital archives emerged as a societal response to the limitations of human memory. The digitization of knowledge through books, manuscripts, social media platforms, online repositories, and collaborative knowledge systems has enabled the creation of a large-scale collective corpus of human knowledge. Recent advances in artificial intelligence, particularly Large Language Models (LLMs), have been made possible through the availability of these extensive digital knowledge repositories and advances in computational architectures.

What prompted us to pursue this extension was a question that emerged from that earlier work and proved difficult to set aside: if writing and libraries are how human cognition escapes its biological limits, what does that say about the limits themselves and about the neural architecture that made such escapes necessary in the first place? The present study is an attempt to follow that question into evolutionary biology and computer architecture simultaneously. In particular, we investigate the role of the genomic bottleneck hypothesis in shaping biological intelligence and explore potential parallels between neural scaling in biological organisms and the scaling of computational architectures. The central argument is not that biological and artificial systems are equivalent, but that they may exhibit common information-processing principles that emerge under different evolutionary constraints.

This paper first examines the importance of low-latency neural responses in the survival of biological organisms, particularly insects, birds, reptiles, and mammals. We then discuss the genomic bottleneck hypothesis and its implications for the emergence of increasingly complex and trainable neural architectures. Subsequently, we examine the evolution of computational systems from low-latency CPUs to high-throughput GPUs, and discuss how these developments enabled modern deep learning systems and Large Language Models. Finally, we consider similarities and differences between biological intelligence, collective human knowledge systems, and contemporary generative artificial intelligence.

## 2. Literature Review

The emergence of intelligence has been investigated across multiple disciplines, including evolutionary biology, neuroscience, cognitive science, and artificial intelligence. Traditional evolutionary theories explain the development of intelligence through natural selection, emphasizing adaptation to environmental pressures and survival advantages. Darwin [3] proposed that cognitive capabilities evolved gradually through the same evolutionary mechanisms responsible for biological diversity.

Recent research has explored the genomic bottleneck hypothesis as a possible explanation for the emergence of trainable neural architectures. Shuvaev et al. [4] argued that the information capacity of the genome is insufficient to explicitly encode the detailed neural connectivity observed in complex organisms. Consequently, evolution may have favored compact developmental rules that generate adaptable neural systems capable of learning from environmental interactions. Similarly, Zador [5] questioned purely data-driven learning approaches and suggested that biological neural systems benefit from innate structures shaped through evolutionary processes.

Studies in neuroscience have also highlighted the importance of neural scaling in the development of higher cognitive functions. Research on insects, vertebrates, and mammals demonstrates that increasingly complex neural systems can emerge through the expansion and specialization of neural architectures. Despite possessing relatively small numbers of neurons, insects exhibit sophisticated behaviors including navigation, communication, and environmental adaptation, suggesting that intelligence depends not only on neuron count but also on organizational efficiency. The complete connectome of Caenorhabditis elegans — comprising 302 neurons and approximately 7,000 synaptic connections — has provided foundational evidence that even minimal neural systems can encode complex behavioral repertoires [6, 7].

Parallel developments have occurred in computer science. Early computing architectures focused primarily on low-latency processing through centralized CPUs. As computational demands increased, GPUs emerged as highly parallel architectures optimized for large-scale data processing. The success of deep learning systems and transformer-based architectures [8] has demonstrated the importance of computational scaling and parallel processing in the development of artificial intelligence. Critically, empirical research on neural scaling laws has demonstrated that the performance of large language models improves predictably as a function of model size, dataset size, and compute budget [9, 10], providing a quantitative framework that parallels the scaling dynamics observed in biological neural systems.

Although substantial research exists on biological evolution, neural scaling, and computational intelligence independently, relatively few studies have examined the conceptual relationships between genomic bottlenecks, neural evolution, computational scaling, and large language models within a unified framework. The present study seeks to address this gap by proposing an interdisciplinary perspective connecting these domains.

Table LR1 provides a structured comparison of the key studies that inform this paper, highlighting their focus, method, principal findings, and the gaps that the present study addresses.

**Table LR1**

*Summary and Comparison of Key Studies in the Literature*

| Author(s) / Year | Domain | Focus | Method / Approach | Key Finding | Limitation / Gap | How it differs from the present work |
|---|---|---|---|---|---|---|
| Darwin [3] | Evolutionary biology | Evolution of cognition and behaviour | Theoretical / philosophical | Cognitive capabilities evolved gradually through natural selection; complexity of ant behaviour noted as remarkable relative to neural size | Does not address information-encoding limits of genetic material or computational analogies | Darwin examined the evolution of cognitive capabilities and behaviour through natural selection. This paper considers genomic information constraints and compares biological neural systems with developments in computational hardware. |
| White et al. [6] | Neuroscience / connectomics | Complete connectome of C. elegans | Empirical — electron microscopy mapping | First full mapping of a nervous system: 302 neurons, ~7,000 synapses; rich behavioural repertoire from minimal circuitry | Single species; cannot generalise scaling principles across organisms without comparative data | White et al. mapped the C. elegans nervous system, while this paper cites their work as background for its discussion of biological neural systems. |
| LeCun, Bengio & Hinton [11] | Artificial intelligence / deep learning | Deep learning architectures and capabilities | Review — computational / empirical | Demonstrated that multi-layer neural networks trained on large datasets achieve human-level | Does not address genomic or evolutionary constraints on biological learning systems | Their review describes advances in deep-learning architectures. By contrast, this paper focuses on |

| Author(s) / Year | Domain | Focus | Method / Approach | Key Finding | Limitation / Gap | How it differs from the present work |
|---|---|---|---|---|---|---|
| | | | | performance on perception tasks | | information constraints in biological systems and discusses their relationship to computational hardware development. |
| Zador [5] | Computational neuroscience / AI | Inductive biases in biological vs. artificial neural networks | Theoretical critique | Argued that biological systems rely on innate genomically encoded structure rather than pure data-driven learning; proposed rethinking AI architectures accordingly | Theoretical argument; empirical tests of GBH-inspired AI architectures remain limited | Zador connected genomic encoding and biological inductive biases with implications for AI. Building on this discussion, this paper considers the potential value of architectural biases in artificial systems facing data constraints. |
| Kaplan et al. [9] | Artificial intelligence / language modelling | Scaling laws for large language models | Empirical — large-scale training experiments | Performance of LLMs improves predictably as a power-law function of model size, dataset size, and compute; identified compute-optimal training regimes | Focused on English-language transformer models; scaling behaviour in other architectures or data regimes not fully characterised | Kaplan et al. established empirical scaling relationships through LLM training experiments. This paper draws on those findings in its theoretical comparison of computational scaling and biological information constraints. |
| Shuvaev et al. [4] | Computational neuroscience / theoretical biology | Genomic bottleneck hypothesis — information encoding in synapses | Theoretical / computational modelling | Genomic capacity falls far short of what explicit specification of synaptic connectivity would require; evolution favours compact developmental programs that generate adaptable, trainable architectures | Theoretical model; direct empirical quantification of genomic information per synapse remains methodologically challenging | Shuvaev et al. formulated the genomic bottleneck hypothesis in relation to synaptic information encoding. This paper applies the hypothesis's predictions to its conceptual comparison of computational scaling and hardware development. |
| Varshney et al. [7] | Computational neuroscience / network science | Structural properties of the C. elegans neuronal network | Empirical — network analysis of connectome data | Analysed the structural properties of the C. elegans neuronal network using connectome data, providing insights into | Limited to a single invertebrate model organism with the simplest known nervous system | Varshney et al. analysed the structural properties of the C. elegans neuronal network. This paper uses their findings as |

| Author(s) / Year | Domain | Focus | Method / Approach | Key Finding | Limitation / Gap | How it differs from the present work |
|---|---|---|---|---|---|---|
| | | | | the organisation of a simple nervous system. | | biological context for its broader discussion of genomic constraints and neural scaling. |
| Vaswani et al. [8] | Artificial intelligence / NLP | Transformer architecture for sequence modelling | Empirical — computational | Introduced attention-based architecture that replaced recurrent networks; enabled highly scalable parallel training of language models | Architecture motivated by engineering efficiency, not biological plausibility; no genomic or evolutionary framing | Vaswani et al. introduced the Transformer architecture for sequence modelling. This paper refers to that architecture when discussing the development of LLMs and external knowledge systems. |
| Hoffmann et al. [10] | Artificial intelligence / language model scaling | Compute-optimal scaling (Chinchilla) | Empirical — large-scale training experiments | Showed that prior LLMs were significantly undertrained relative to model size; optimal scaling requires proportional growth of model size and data | Findings specific to autoregressive transformer LLMs; broader generalisability to other architectures or biological systems not tested | Hoffmann et al. experimentally examined compute-optimal LLM training and model–data scaling, whereas this paper uses their findings in a conceptual comparison of computational scaling and biological information constraints in Section 9. |
| Clark & Chalmers [1] | Cognitive science / philosophy of mind | Extended mind thesis — cognitive boundaries | Philosophical / theoretical | Argued that cognitive processes extend beyond the brain into tools, artefacts, and social systems; external objects can constitute genuine parts of cognitive systems | Philosophical argument; empirical operationalisation of extended cognition boundaries remains debated in cognitive science | Clark and Chalmers proposed the extended mind thesis. This paper connects that thesis to digital repositories and LLMs to discuss the externalisation of human knowledge and cognition. |

*Note.* GBH = Genomic Bottleneck Hypothesis; LLM = Large Language Model. Studies selected on the basis of direct relevance to the paper's core argument. Method/Approach column reflects the primary research strategy of each work.

## 3. The Genomic Bottleneck Hypothesis

The genomic bottleneck hypothesis (GBH) offers a fundamental explanation for how complex, adaptive neural architectures can arise from the comparatively limited information storage capacity of the genome. The human genome contains approximately 3.2 billion base pairs and encodes roughly 20,000 protein-coding genes. Yet the adult human brain contains approximate 86 billion neurons forming approximately 100 trillion synaptic connections. The information required to specify these connections explicitly would far exceed the storage capacity of the genome by several orders of magnitude. This disparity constitutes the genomic bottleneck: hereditary information is fundamentally insufficient to pre-wire a fully specified nervous system.

Shuvaev et al. [4] formalized this constraint by showing that the information capacity of the genome is much lower than would be needed for an explicit specification of connectivity. They proposed that evolution instead favors compact developmental programs — algorithmic rules that generate connectivity patterns through growth, activity-dependent refinement, and experience-dependent plasticity. Under this view, the genome does not specify the brain; it specifies a developmental process that builds a brain capable of learning from environmental experience.

Zador [5] extended this argument by critiquing purely data-driven approaches to artificial intelligence. He noted that biological neural systems arrive pre-equipped with strong inductive biases — architectural priors that constrain and guide learning — rather than learning everything from scratch. These innate biases reflect compressed evolutionary knowledge encoded through the genomic bottleneck. The implication for artificial systems is significant: systems facing data scarcity during training may benefit from architecturally encoding inductive biases rather than relying solely on statistical learning from large corpora.

The GBH generates several testable predictions relevant to this study. First, organisms subject to stronger genomic constraints should exhibit higher degrees of postnatal plasticity and environmental sensitivity, as learning must compensate for what cannot be pre-specified. Second, neural architectures that emerge under genomic constraints should exhibit efficient, reusable computational modules rather than hard-coded, task-specific circuits. Third, the transition from genetically determined to experientially shaped intelligence should correlate with increases in genomic constraint relative to neural complexity — a pattern that Table 1 illustrates across species ranging from C. elegans to Apis mellifera. These predictions connect directly to the computational analogies developed in subsequent sections and motivate the comparison between biological neural scaling and the scaling behavior of large language models [9, 10].

## 4. Research Questions and Objectives

This study seeks to explore the relationship between biological neural evolution and the evolution of computational intelligence systems. Specifically, the study addresses the following research questions:

> RQ1: How does the genomic bottleneck hypothesis contribute to our understanding of neural scaling in biological organisms?
>
> RQ2: What conceptual similarities exist between biological neural systems and modern computational architectures?
>
> RQ3: How can the development of large language models be interpreted through the framework of biological information processing and neural scaling?

The primary objective of this study is to develop an interdisciplinary theoretical framework connecting evolutionary biology, neuroscience, computer architecture, and artificial intelligence. Through this synthesis, the study aims to provide new perspectives on the emergence of intelligence in both natural and artificial systems.

## 5. Methodology

This study adopts a theoretical synthesis approach, drawing on an interdisciplinary narrative review of literature from evolutionary biology, neuroscience, computer architecture, and artificial intelligence. Rather than generating primary empirical data, the study constructs a conceptual framework by identifying common information-processing principles across biological and computational domains.

The literature search was conducted across multiple academic databases including PubMed, Google Scholar, IEEE Xplore, and arXiv. Search terms included: genomic bottleneck, neural scaling, insect neuroscience, C. elegans connectome, CPU evolution, GPU architecture, deep learning scaling laws, large language models, and extended cognition. Sources were selected based on their relevance to the core argument, empirical grounding, and citation impact. Priority was given to foundational empirical studies, peer-reviewed journal articles, and preprints from recognized research groups. Hardware specification data were compiled from manufacturer documentation and published technical sources. Representative CPU and GPU models were selected to illustrate changes in processing architecture, core counts and memory bandwidth over time. The specifications are used for historical comparison and are not intended to represent a complete dataset of processor performance.

The conceptual mapping between biological and computational systems was constructed through a process of analogical reasoning. Analogies were evaluated against two criteria: (1) structural correspondence — whether the systems share functionally similar constraints or organizational principles — and (2) generativity — whether the analogy produces insights or predictions that neither field could derive independently. Where analogies are illustrative rather than explanatory, this is explicitly noted. The study does not claim that biological and digital systems are equivalent; it claims that shared information-processing challenges produce structurally similar solutions across evolutionary and engineering timescales.

## 6. Low-Latency Neural Systems and Biological Survival

The survival of biological organisms often depends upon the ability to perceive environmental changes and respond within extremely short time intervals. Across evolutionary history, insects, birds, reptiles, and mammals have developed neural systems optimized for rapid sensory processing and behavioral responses. These systems provide immediate advantages in predator avoidance, hunting, navigation, reproduction, and environmental adaptation.

Insects provide particularly interesting examples of low-latency neural architectures. Despite possessing relatively small nervous systems, many insects demonstrate remarkable behavioral sophistication, including complex flight control, navigation, communication, and environmental adaptation. The survival of flying insects such as mosquitoes and flies depends heavily on their ability to respond rapidly to predators and environmental disturbances. Such capabilities suggest that evolutionary pressures have favored neural architectures optimized for speed, efficiency, and reliability rather than for extensive postnatal learning.

Although many insect behaviors appear largely innate, learning is not entirely absent. Studies of social insects such as honey bees demonstrate that learning can play an important role in communication and resource acquisition [12]. Nevertheless, compared with larger vertebrates, many insects rely more heavily on genetically encoded behavioral programs and shorter developmental cycles, reflecting the evolutionary pressures associated with high reproduction rates and relatively short lifespans.

**7. Insect Neural Architectures and Computational Analogies**

One of the most intriguing aspects of insect nervous systems is their ability to generate highly sophisticated behavior using relatively small numbers of neurons. Charles Darwin observed that the behavioral complexity of ants is particularly remarkable given the extremely small size of their nervous systems. This observation continues to inspire contemporary discussions regarding efficiency in biological information processing.

Unlike contemporary Large Language Models, insect neural systems do not merely process information passively. They continuously integrate sensory inputs with motor outputs, enabling real-time interaction with dynamic environments. Their neural networks function simultaneously as perception, decision-making, and action systems.

Some insects are capable of maintaining limited behavioral functions even after severe damage to central neural structures. While such observations should not be interpreted as evidence that insects lack brains, they do suggest a degree of distributed processing within their nervous systems. This distributed organization contrasts with the stronger centralization observed in many vertebrate nervous systems.

To provide an intuitive perspective on neural complexity, it is useful to draw illustrative comparisons with computational architectures. Such comparisons should not be interpreted as direct equivalences between biological and digital systems. Biological neurons, synapses, and developmental processes differ fundamentally from digital processors. Nevertheless, both systems face common challenges related to information processing, memory constraints, response speed, and resource optimization.

A particularly instructive case is Caenorhabditis elegans, one of the first organisms for which a complete nervous-system connectome was mapped, providing an important reference for understanding the relationship between neural connectivity and biological information processing [6, 7]. With precisely 302 neurons and approximately 7,000 synaptic connections, executes a rich repertoire of behaviors including chemotaxis, thermotaxis, mechanosensation, and associative learning. The fact that such behavioral complexity can arise from so few neurons and such a compact genome (approximately 100 million base pairs) illustrates the GBH mechanism directly: the genome encodes developmental rules, not explicit connectivity, enabling flexible behavior from minimal genetic specification.

Under this interpretation, the neural complexity of some insects may be viewed as loosely analogous to the computational capabilities of early microprocessor architectures. The purpose of this analogy is not to establish computational equivalence but to highlight how relatively small systems can produce surprisingly sophisticated behavior through efficient organization and optimization.

To provide an illustrative comparison of neural complexity across selected species, Table 1 summarizes the approximate number of neurons, genome sizes, reproductive characteristics, lifespan, and physical dimensions of several insects and other small organisms.

**Table 1**

*Neural and Genomic Characteristics of Selected Invertebrate Species*

| Species | Neurons | Base Pairs | Parenting | Size | Lifespan | No. of eggs |
|---|---|---|---|---|---|---|
| Culex quinquefasciatus (Mosquito) | 225,911 ± 7,220 (approximately 226,000) | 1.38 billion | None | 5 mm | 14 days | 1,200 |
| Tribolium ferrugineum (Rusty grain beetle) | ~100,000 | ~165 million | Eggs in stored grain | 2–3 mm | 1–2 years | Up to 400 |
| Cimex lectularius (Bed bug) | ~250,000 | ~700 million | Eggs near host animals | 4–5 mm | 6–12 months | Up to 500 |
| Musca domestica (House fly) | ~100,000 | ~1.5 billion | Eggs in organic matter | 6–9 mm | 15–30 days | Up to 500 |
| Tribolium confusum (Confused flour beetle) | ~100,000 | ~160 million | Eggs in stored food | 3–4 mm | 1–2 years | Up to 400 |
| Tribolium castaneum (Red flour beetle) | ~100,000 | ~163 million | Eggs in flour sources | 2–3 mm | 2–12 months | Up to 500 |
| Apis mellifera (Honey bee) | ~960,000 | ~250 million | Social; queen lays eggs | 12–15 mm | 5 wks (worker); yrs (queen) | 2,000/day (queen) |
| Caenorhabditis elegans (Roundworm) | 302 | ~100 million | Self-fertilization | 1 mm | 2–3 weeks | ~300 |
| Drosophila melanogaster (Fruit fly) | ~100,000 | ~165 million | None (lay eggs) | 2–3 mm | 40–50 days | Hundreds |

*Note.* Source: Genome sizes are drawn from Adams et al. [13], Benoit et al. [14], He et al. [15], Richards et al. [16], and Wallberg et al. [17]; the C. elegans nervous system from White et al. [6]. Life-history and behavioral data are drawn from Barrière and Félix [18], Delaplane and Mayer [19], Doggett et al. [20], Félix and Braendle [21], Félix and Duveau [22], Greenberg [23], Partridge and Fowler [24], Reinhardt and Siva-Jothy [25], Simon et al. [26], Su et al. [27], Technau and Heisenberg [28], and Winston [29]. Neurons and base pair values are approximate.

Table 1 demonstrates that complex behavior can emerge from relatively small neural systems. Species such as honey bees and fruit flies exhibit sophisticated navigation, communication, and adaptive behaviors despite possessing far fewer neurons than vertebrate species. This observation supports the argument that intelligence and behavioral complexity depend not only on the number of neurons but also on the efficiency of neural organization and information processing mechanisms — a pattern consistent with the predictions of the genomic bottleneck hypothesis outlined in Section 3.

The historical growth of computational performance has been strongly influenced by Moore's Law, which describes the exponential increase in transistor density over time. This trend enabled significant improvements in processor performance, memory capacity, and computational capability. Figure 1 illustrates the long-term growth trajectory of computing systems associated with Moore's Law.

*Figure 1. Moore's Law and the Evolution of CPU Performance*

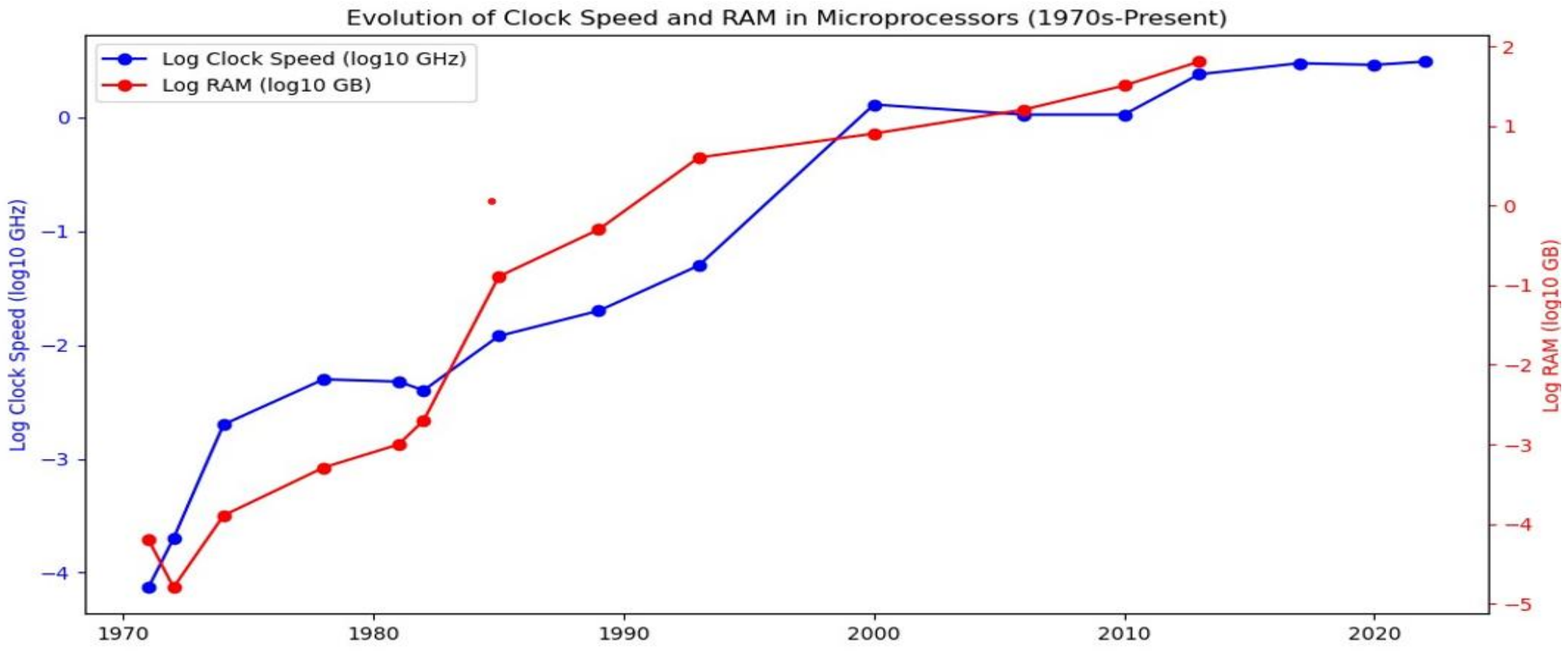

*Source: Adapted from Moore's Law historical data.*

Figure 1 illustrates the sustained growth in computational capability over several decades.

The continuous increase in transistor density enabled increasingly complex software systems and laid the technological foundation for modern artificial intelligence applications. However, physical limitations in transistor miniaturization eventually encouraged the development of alternative approaches such as multi-core architectures and GPU-based parallel processing.

The development of central processing units (CPUs) provides an important technological parallel to the evolution of biological neural systems. Early processors were primarily optimized for low-latency sequential processing, similar to the rapid response requirements observed in many biological organisms. Table 2 summarizes the evolution of representative CPU architectures from 1974 to 2022.

**Table 2**

*Evolution of Intel Microprocessor Clock Speeds and Cache Architecture (1974–2022)*

| Year | Microprocessor | Clock Speed (GHz) | L1 Cache (KB) | L2 Cache (MB) |
|---|---|---|---|---|
| 1974 | Intel 8080 | 0.002 | — | — |
| 1978 | Intel 8086 | 0.005–0.01 | — | — |
| 1981 | Intel 8088 | 0.00477 | — | — |
| 1985 | Intel 386 | 0.012–0.033 | — | — |
| 1989 | Intel 486 | 0.02–0.1 | 8 | — |
| 1993 | Intel Pentium | 0.05–0.3 | 16 | — |
| 2000 | Intel Pentium 4 | 1.3–3.8 | 8–16 | 0.256–2 |
| 2006 | Intel Core 2 Duo | 1.06–2.66 | 32–64 | 2–6 |
| 2010 | Intel Core i7 | 1.06–3.6 | 64–128 | 4–12 |
| 2017 | Intel Core i9 | 3.0–4.3 | 256–512 | 8–24 |
| 2020 | Intel Core i9 | 2.9–5.3 | 512–1024 | 12–24 |
| 2022 | Intel Core i9 | 3.1–5.3 | 512–2048 | 16–32 |

*Note. Source: Compiled from Intel, AMD, IBM, and historical processor documentation. Table shows representative Intel milestones; full dataset available from authors. Cache capacities are approximate and vary across processor models and configurations within each generation. In multi-core processors, reported L1 cache capacity may represent either per-core or aggregate capacity. Therefore, the values are intended to illustrate the broad evolution of cache architecture rather than provide a direct performance comparison between processors.*

Table 2 shows the significant growth in computational performance achieved through different generations of processor development. Three key observations are relevant to the paper's argument, each revealing a unique aspect of the biological-computational parallel proposed by the GBH framework.

First, the shift from kiloFLOPS performance in the 1970s (Intel 8080: ~0.002 MFLOPS) to hundreds of GFLOPS in the 2020s (Core i9: 677-861 GFLOPS) represents about an eight-order-of-magnitude improvement in raw computational power over five decades. This trend parallels the increase in neural complexity over evolutionary time, from the 302-neuron nervous system of *C. elegans* to the approximate 86 billion neurons in the human brain. In both cases, the increase was not simply due to adding identical units. In computing, newer processor generations brought different architectural innovations, such as pipelining, superscalar execution, out-of-order processing, and cache hierarchies. In biology, the evolution of the cerebral cortex involved increasingly specialised and layered organisation, columnar processing, and long-range connectivity patterns associated with more complex cognitive functions. The Genomic Bottleneck Hypothesis (GBH) offers one possible theoretical framework for interpreting this development. It suggests that the limited capacity of the genome to encode neural connectivity explicitly may have favoured the evolution of more complex organisational and developmental mechanisms. However, this interpretation should be considered alongside other evolutionary and biological explanations rather than as an established or exclusive account of neural complexity.

Second, the stagnation in single-core clock speed growth visible in Table 2 after about 2004—the ‘power wall’—is structurally similar to the genomic bottleneck in biology. By the mid-2000s, raising transistor clock frequency became unsustainable due to thermodynamic limits: power use and heat output increased faster than performance improvements. The response was not to abandon silicon but to reorganize it by introducing multi-core architectures that achieved higher overall throughput through parallel execution instead of serial. The biological analogy is instructive: the genome faces a similar limit on the ratio of information capacity to synaptic complexity. Evolution responded with a reorganization of architecture, moving from genetically specified circuitry to experience-driven plasticity, allowing the brain to adjust its connections based on environmental inputs as a secondary specification layer. Both transitions fit a general pattern that when a primary scaling approach hits a limit, progress tends to continue via a complementary architectural approach. We are employing this as a structural analog only . The mechanisms are different, as the processor limit is thermal and the biological limit is informational .

Third, the development of cache hierarchies—from none in early processors to several megabytes by 2022—is a computing response to the increasing gap between processor speed and memory access time. Biological systems are organized very differently, so we give the comparison only as an illustration of added organization compensating for a limit on direct access. The fact that these solutions are similar in very different systems provides modest support for the paper’s main argument: A useful framework for thinking about this is the GBH: shared information-processing limits can lead to structurally similar solutions..

## 8. Limitations of Human Cognitive Capacity and the Emergence of External Knowledge Systems

Despite possessing one of the most complex neural architectures known in nature, the human brain remains subject to significant biological constraints. Human memory is inherently selective, lossy, and limited in its ability to preserve large quantities of information across extended periods of time. As human societies increased in complexity, mechanisms for preserving and transmitting knowledge became increasingly important.

Various cultural innovations emerged as responses to these cognitive limitations. Oral traditions, poetry, rhythm, music, and storytelling provided mechanisms for improving memory retention and intergenerational transmission of knowledge. The subsequent development of writing systems enabled information to be preserved independently of biological memory, allowing knowledge to be recorded, transmitted, and accumulated across generations.

The establishment of libraries, archives, and educational institutions further expanded humanity's capacity for collective knowledge storage. The invention of the printing press dramatically increased access to information, accelerating scientific, technological, and cultural development. These external knowledge systems effectively extended human cognitive capabilities beyond the limitations of individual memory and biological information processing.

This perspective aligns with the extended mind thesis [1, 30], which argues that cognitive processes are not bounded by the skull but extend into tools, artifacts, and social systems. Writing, libraries, and digital repositories are not merely aids to cognition; under the extended mind framework, they constitute genuine components of distributed cognitive systems. The evolution of human civilization can therefore be viewed as a progressive externalization of cognitive architecture — a macroscale analog of the genomic bottleneck's solution at the cellular level.

## 9. Computational Scaling: From CPUs to GPUs and Specialized Hardware

As improvements in traditional CPU scaling began to slow, graphics processing units emerged as an alternative architecture optimized for massively parallel computation. This transition has been described by Huang's Law, which suggests that GPU performance improvements have significantly outpaced traditional CPU performance growth. Figure 2 illustrates this trend.

*Figure 2. Huang's Law and GPU Computational Growth*

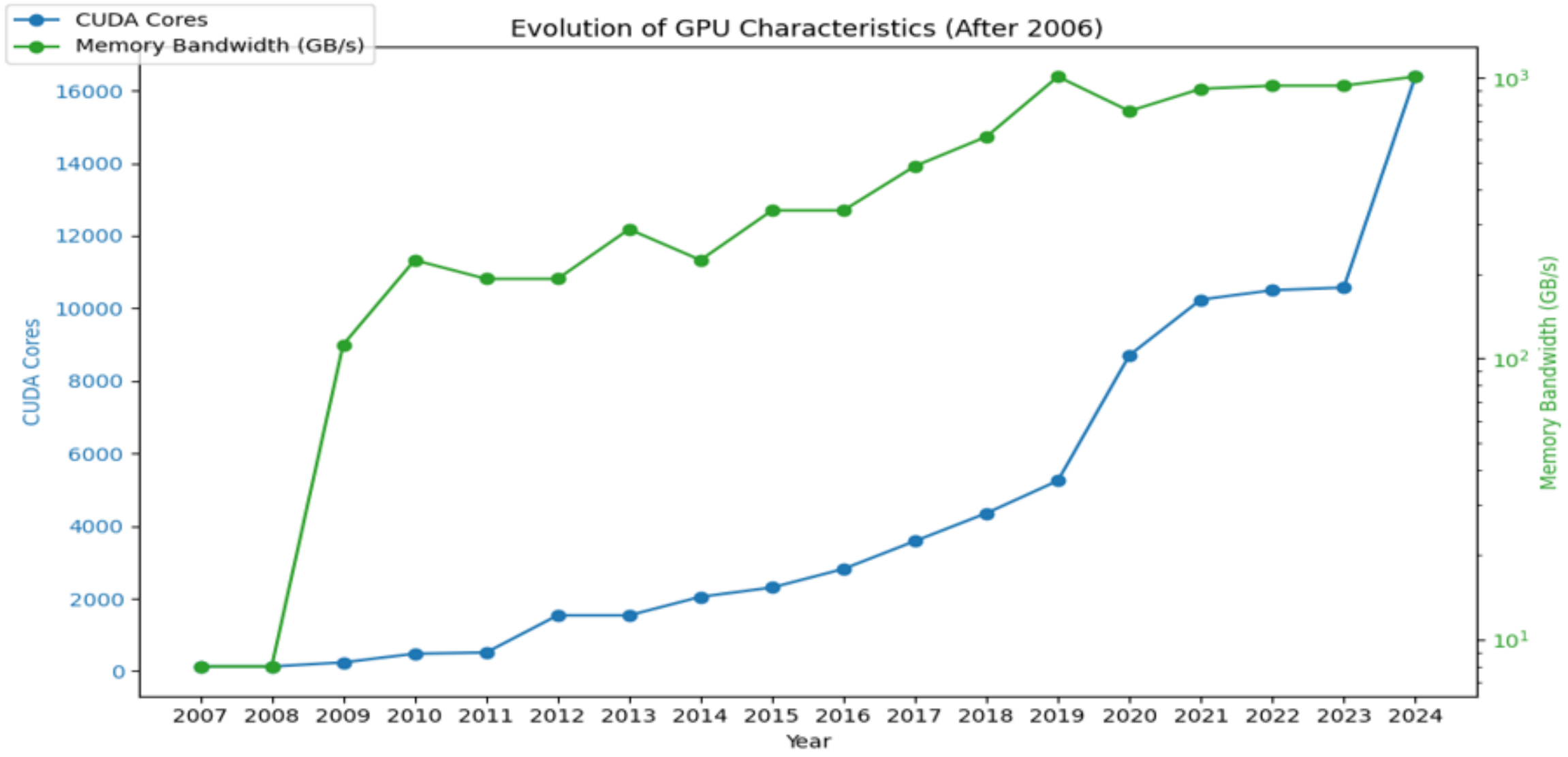

*Source: Adapted from NVIDIA and industry GPU performance trends.*

Figure 2 highlights the rapid increase in GPU computational capability relative to traditional CPU performance growth. The availability of high-throughput parallel processing architectures played a critical role in enabling deep learning systems, transformer architectures, and modern Large Language Models.

The historical development of computing systems can be understood as a sequence of technological transitions from low-latency sequential processing toward increasingly parallel and specialized architectures. Table 3 summarizes major milestones in the evolution of CPUs and GPUs.

**Table 3**

*Stages of CPU and GPU Evolution*

| Era | Milestones | References |
|---|---|---|
| 1970s | Introduction of first microprocessors (Intel 4004, 8008) | [31],[32] |
| 1980s | IBM PC launched with Intel 8088 CPU | [31] |
| 1990s | Introduction of Intel Pentium series CPUs<br>Rise of dedicated 3D graphics cards (e.g., NVIDIA RIVA) | [33]<br>[34] |
| 2000s | Emergence of multi-core CPUs (e.g., Intel Core series)<br>GPUs transition to programmable architectures<br>Introduction of GPGPU computing | [35]<br>[36],[34]<br>[36] |
| 2010s | Deep learning boom drives demand for GPU acceleration<br>Introduction of CUDA and OpenCL for GPU programming<br>Development of specialized AI accelerators (e.g., TPUs) | LeCun et al. [11]<br>[37]<br>[38] |
| Present | Continued development of multi-core CPUs<br>GPUs with dedicated tensor cores for AI workloads | [35]<br>[37] |
| Future | Advancements in heterogeneous computing architectures | [39] |

*Note.* Source: Compiled from Intel, NVIDIA, AMD, and Google TPU documentation.

Table 3 highlights how computational architectures evolved through specific technological eras, each marking a clear shift in the main information-processing model. Seen through the GBH framework, this progression reveals a pattern of constraint-driven architectural innovation that aligns closely with the pattern found in biological neural evolution.

The move from single-core CPUs in the 1970s and 1980s to multi-core architectures in the 2000s was not a smooth transition but a sudden architectural shift forced by physical limits—the power wall mentioned earlier. The rise of programmable GPU architectures and GPGPU computing in the 2000s signified another major change: a shift from serial, low-latency optimization suitable for general tasks to massively parallel optimization suitable for large-scale matrix operations. This transition didn't just happen; it was driven by specific application needs—real-time 3D graphics processing—that created pressure for a fundamentally different architecture. A loosely analogous case in biology might be the enlargement of the visual cortex in primates, where specialized ecological challenges are thought to have favored specialized neural systems.

The rise of specialized AI accelerators (TPUs, 2010s) marks a third pivotal change: a move from general-purpose computing to domain-specific optimization for machine learning tasks. This is a crude analog for the specialization observed in neural circuits such as the insect mushroom bodies, the mammalian hippocampus and the primate prefrontal cortex. The GBH offers a possible interpretation of this trend: when generalized plasticity reaches its limits as a learning strategy, evolution may favor genetically specified architectural modules that prepare the learning system for recurrent ecological challenges. Table 3 is more than a history of technology in this reading. It can also be viewed as an example of problem solving under constraints, although we do not claim the processes to be the same.

The rapid evolution of Graphics Processing Units (GPUs) has been a major factor enabling the development of large-scale artificial intelligence systems. Unlike traditional CPUs, which prioritize low-latency sequential processing, GPUs are optimized for massively parallel computation. Table 4 illustrates the growth of NVIDIA GPU architectures from 2008 to 2022, presented chronologically.

**Table 4**

*Evolution of NVIDIA GPU Architectures and Processing Capabilities (2008–2022)*

| Year | GPU Model | CUDA Cores / Stream Processors | Memory Interface (bit) | Memory Bandwidth (GB/s) |
|---|---|---|---|---|
| 2008 | GeForce GTX 260 (192-core version) | 192 | 448 | 111.9 |
| 2008 | GTX 280 | 240 | 512 | 141.7 |
| 2009 | GTX 295 (dual-GPU) | 480 | 896 | 223.8 |
| 2010 | GTX 480 | 480 | 384 | 177.4 |
| 2010 | GTX 580 | 512 | 384 | 192 |
| 2012 | GTX 690 (dual-GPU) | 3,072 | 512 (256 per GPU) | 384 |
| 2013 | GTX 780 Ti | 2,880 | 384 | 336 |
| 2014 | GTX 980 | 2,048 | 256 | 224 |
| 2015 | GTX 980 Ti | 2,816 | 384 | 336.5 |
| 2016 | GTX 1080 | 2,560 | 256 | 320 |
| 2017 | GTX 1080 Ti | 3,584 | 352 | 484 |
| 2018 | RTX 2080 | 2,944 | 256 | 448 |
| 2018 | RTX 2080 Ti | 4,352 | 352 | 616 |
| 2020 | RTX 3080 (10 GB) | 8,704 | 320 | 760 |
| 2020 | RTX 3090 | 10,496 | 384 | 936 |
| 2021 | RTX 3080 Ti | 10,240 | 384 | 912 |
| 2022 | RTX 3090 Ti | 10,752 | 384 | 1,008 |
| 2022 | RTX 4090 | 16,384 | 384 | 1,008 |

*Note.* Source: Specifications are adapted from the NVIDIA Graphics Card Specification Chart (Studio 1 Productions; David Knarr) and cross-checked against available NVIDIA product specifications and architecture documentation. CUDA core counts, memory-interface widths and memory bandwidth are reported for the respective GPU configurations. The GTX 690 is a dual-GPU card; its CUDA core count and bandwidth represent the combined configuration.

Table 4 captures one of the most significant hardware scaling journeys in computing history. Between 2008 and 2022, CUDA core counts jumped from 192 to 16,384—while memory bandwidth soared from around 112 GB/s to over 1,000 GB/s. This growth wasn't linear: Table 4 shows two notable changes in the reported core counts, the rise from ~500 to ~3,000 CUDA cores between the GTX 580 (2010) and the GTX 690 (2012, a dual-GPU card), and the more dramatic leap from ~4,352 to ~8,704 cores between the RTX 2080 Ti (2018) and RTX 3080 (2020). These changes can be viewed as loosely analogous to the pattern of punctuated equilibrium, that is described in biological evolution, where long periods of gradual change are interrupted by swift shifts made possible by new developmental or ecological chances.

The GPU scaling path is crucial for understanding the rise of large language models because it establishes the computational infrastructure necessary for large-scale transformers. For instance, the RTX 3090's 936 GB/s memory bandwidth is about 5.3 times in terms of memory bandwidth than the GTX 480's 177.4 GB/s from 2010. This improvement in bandwidth helps support the large matrix multiplications that form the core functions of transformer attention mechanisms. Kaplan et al. [9] showed that LLM performance scales as a power law with compute; Hoffmann et al. [10] found that optimal scaling requires proportional growth in both model parameters and training data. The GPU scaling that is outlined in Table 4 contributed to the computational infrastructure for training at the scales these laws describe. Continued hardware improvements therefore helped make such training scales increasingly practical.

From the perspective of the GBH framework, the GPU scaling journey offers an analogy for how artificial systems address constraints in information representation and learning. Just as biological evolution does not explicitly encode every connection in the adult connectome within the genome and thus evolved systems that learn connections through experience, GPU-accelerated LLMs do not rely on explicitly programmed knowledge representations and instead learn from data. The hardware infrastructure detailed in Table 4 helped support that data-driven learning at the scale needed to create emergent abilities. In this sense, GPU scaling plays a supporting role in artificial systems, just as evolutionary time provides a broader context for biological systems: it provided the computational foundation within which complex, learned structures could develop.

## 10. Social Extension of Human Intelligence: From Written Knowledge to Large Language Models

The limitations of biological memory encouraged the development of increasingly sophisticated external information storage systems. Writing, books, libraries, and archives enabled knowledge to be preserved beyond individual lifespans. The emergence of digital technologies further accelerated this process through searchable databases, online repositories, and collaborative knowledge platforms such as Wikipedia. This progression constitutes what Clark and Chalmers [1] describe as the extended mind: cognitive processes that span individual brains and their external scaffolding.

At the same time, advances in computer architecture transformed the ability of machines to process information. While traditional CPUs were primarily optimized for low-latency sequential processing, GPUs introduced massively parallel computation capable of supporting large-scale matrix operations. These developments proved particularly important for deep learning systems and transformer-based architectures. The transformer architecture introduced by Vaswani et al. [8] enabled highly scalable attention-based neural networks and laid the foundation for contemporary Large Language Models.

Large Language Models can therefore be interpreted as an extension of humanity's long-standing effort to externalize, organize, and retrieve knowledge. Trained on vast digital corpora generated through collective human activity, these systems exhibit capabilities that resemble certain aspects of human cognition, including language generation, knowledge retrieval, and pattern recognition. The scaling behavior of these systems — in which performance has been observed to improve as a power law with model size, data, and compute [9] and where compute-optimal training requires proportional scaling of both model size and dataset size [10] — can be considered alongside the broader scaling relationships observed in biological neural systems, although the underlying mechanisms differ. In some settings, increasing the scale of a learning system has been associated with the appearance of qualitatively new capabilities, although the nature and measurement of these emergent abilities remain subjects of ongoing research

However, they also exhibit limitations such as bias, hallucination, and memorization, reflecting characteristics present within the underlying data from which they learn. These limitations parallel characteristics of human collective intelligence: societies similarly propagate biases, historical inaccuracies, and misinformation through their shared knowledge systems. Understanding these parallels may provide insights for developing more reliable and socially aligned artificial intelligence.

## 11. Conclusion

This study explored the relationship between the genomic bottleneck hypothesis, neural scaling, and the evolution of computational architectures as a framework for understanding the emergence of intelligence in both biological and artificial systems. The analysis addresses the three research questions posed in Section 4.

Regarding RQ1, the genomic bottleneck hypothesis contributes to our understanding of neural scaling by explaining why increasingly complex organisms exhibit increasingly plastic and trainable neural architectures. Given the limits of genomic encoding, one possible interpretation is that evolution has favoured compact developmental programs that generate adaptable systems capable of learning from environmental experience. The GBH suggests that information constraints may be relevant to understanding the development and organisation of biological neural systems. However, the available data does not establish a direct causal relationship between genomic information constraints and neural complexity across species.

Regarding RQ2, biological neural systems and modern computational architectures exhibit several conceptual parallels rooted in shared information-processing constraints. Both domains encounter constraints on system development and computational capacity — the genomic bottleneck in biology, and the power wall and memory bandwidth bottleneck in computing. Both responded to these limits through architectural shifts: cortical expansion and social cognition in biology; multi-core processing and GPU-based parallelism in computing. The analogy provides a theoretical framework for considering whether artificial systems facing training-data scarcity may benefit from architecturally encoded inductive biases, analogous in a broad structural sense to the innate structures produced by genomic bottleneck pressures in biological systems. This proposition requires experimental validation.

Regarding RQ3, the development of large language models can be interpreted as the latest stage in humanity's long-term effort to externalize and extend cognition beyond biological constraints. Enabled by GPU-scale parallel computation and trained on collectively produced digital corpora, LLMs represent a technological continuation of the extended mind process that began with writing and libraries. Their scaling behavior — scaling relationships between model performance, model size, training data, and computational resources [9, 10] — can be considered alongside the broader scaling relationships observed in biological neural systems, although the underlying mechanisms differ, suggesting that common information-processing principles may underlie intelligence emergence across both domains.

The limitations of biological cognition have historically been supplemented through external knowledge systems such as language, writing, libraries, and digital information repositories. Contemporary artificial intelligence systems may be viewed as a continuation of this process. The transition from low-latency CPU-centric computing to massively parallel GPU-based computation provides an interesting technological parallel to the scaling processes observed in biological neural systems. While such comparisons should be interpreted as conceptual analogies rather than direct equivalences, they offer a useful perspective for understanding how increasing complexity and scale can produce new forms of information processing capability.

Many aspects of biological intelligence, social learning, and collective human behavior remain only partially understood and have yet to be fully reflected in contemporary artificial neural architectures. Nevertheless, several limitations observed in large language models — including bias, memorization, hallucination, and the propagation of misinformation — display intriguing parallels with characteristics of human societies and collective intelligence. Understanding these similarities may provide valuable insights for the development of more reliable, interpretable, and socially aligned artificial intelligence systems.

Although the framework proposed in this study remains theoretical, it offers a potential foundation for future interdisciplinary investigations into the evolution of intelligence across biological and artificial domains.

**12. Contributions of the Study**

This study brings together concepts from evolutionary biology, neuroscience and machine learning to examine how information constraints may have influenced the development of complex information-processing systems. The connections we discussed here should not be understood as direct equivalence between biological and computational processes. Instead, our study develops three related propositions based on previous literature.

First, examining the genomic bottleneck hypothesis (GBH) in relation to computational scaling.

The role of information constraints in the development of biological systems has been discussed in previous studies [4, 5]. These studies examine how complex neural systems can develop despite the limited information available in the genome. The present paper extends this discussion by considering whether a similar pattern of constraint-driven change can be observed in computing architecture. The transition from single-core to multi-core processing, in response to the limitations associated with the CPU power wall, provides a useful point of comparison. In both cases, simply increasing capacity is not sufficient to address the constraints involved; changes in architecture also become important. However, the nature of these constraints differs: biological systems face limitations in the genomic information available for development, whereas computing systems face physical, power and thermal constraints. This comparison provides a basis for examining how different information-processing systems respond to limitations through architectural change.

Second, considering whether biological inductive biases may inform artificial intelligence under data constraints.

Zador [5] argued that biological neural systems benefit from structural inductive biases shaped through evolution. These provide an initial framework for learning, rather than requiring biological systems to acquire all the necessary structure from large amounts of data. This raises an important question for artificial intelligence: could artificial systems also benefit from suitable architectural biases when training data are limited? The present study develops this question as a theoretical proposition. It suggests that artificial systems incorporating appropriate prior structural assumptions may benefit in few-shot and low-data learning settings. This proposition has not been tested experimentally in the present study. Instead, it identifies a direction for future research, in which different architectural choices could be compared under limited training-data conditions to examine their influence on generalisation.

Third, relating the extended mind thesis [1] to the GBH to examine cognition across three levels.

At the cellular level, the discussion focuses on how information stored in the genome contributes to the development of neural connectivity. At the organismal level, brain architecture and experience contribute to the development of cognitive capabilities. At the civilisational level, language, writing, libraries, digital repositories and artificial intelligence enable knowledge to be stored, shared and accessed beyond the memory of an individual. Considering these three levels together provides a broader perspective on how information is organised and made available to support increasingly complex forms of cognition. The extended mind thesis is used here to explore the relationship between biological development and the growth of external knowledge systems. This connection does not, however, imply that biological development and external knowledge systems operate through identical mechanisms.

## 13. Limitations

This study is conceptual and theoretical in nature. The proposed relationships are derived from interdisciplinary synthesis rather than direct experimental validation. The analogies between biological and computational systems, while generative, remain analogies; direct empirical equivalence between genomic bottleneck mechanisms and AI training dynamics has not been established.

Underlying assumptions

The broader goal of this study is it uses approximate genome-size and neuron-count values for nine selected species (Table 1) to illustrate what are the possible relationships between genomic constraints and neural complexity. Both Genome size and neuron count are descriptive measures; they are not the direct measurements of genomic information or neural connectivity. We are treating the genomic bottleneck hypothesis (GBH) as a hypothesis rather than an established fact. Similarly, the selected GPU models and specifications that are illustrated in Table 4 changes in hardware architecture and capacity; they do not provide any direct measures of overall computational performance. The comparisons between biological and computational systems rather than mechanistic they are structural. Although its interpretation remains debated,the extended mind thesis is used as a theoretical framework.

Alternative explanations

The Differences that are observed in the neuron counts across species may also be associated with body size, lifespan, energy requirements and evolutionary adaptations. The Changes in CPU and GPU specifications across Tables 2–4 may reflect manufacturing advances, market demand, gaming and software requirements. The development of large language models has also been influenced by data availability, algorithms and model architecture, alongside improvements in computing hardware. These factors were not independently tested or excluded in this study. Hence, information constraints should be understood as one possible way of perspective for interpreting the developments discussed, rather than as their sole explanation.

Future empirical studies are required to evaluate the proposed framework. In particular, quantitative studies examining whether architecturally encoded inductive biases improve performance under data-scarce training conditions would provide a direct test of the central prediction. Similarly, comparative analyses of connectome data across species with varying genome-to-synapse ratios could provide biological evidence for the GBH scaling predictions outlined in Section 3.

## 14. Future Research Directions

Future studies should investigate the proposed relationships through biological datasets, computational simulations, and analyses of large-scale neural architectures. Such studies may help clarify the common principles underlying intelligence in both biological and artificial systems. Specific directions include: (1) empirical analysis of the relationships between genome size, connectome complexity, and behavioral plasticity across a broader range of species, while accounting for the limitations of genome size as a proxy for genomic information; (2) computational experiments testing whether GBH-inspired architectural priors improve few-shot generalization in artificial neural networks; (3) extension of the extended mind framework to characterize LLMs as components of distributed cognitive systems; and (4) investigation of whether the bias and hallucination characteristics of LLMs can be better understood through the lens of collective human information propagation dynamics.

**Ethics Statement**

This study is conceptual and theoretical in nature and is based on previously published literature and publicly available information. It did not involve human participants, animal experiments, or the collection of personal or identifiable data. Therefore, ethical approval was not required.

**Funding Statement**

This research received no external funding.

**Author Contributions**

Dr. Arifa Khan: Literature review, investigation, formal analysis, data curation, visualization, and writing—original draft preparation.

Dr. Suki Venkatesan: Conceptualization, development of the theoretical framework, methodology, intellectual guidance, critical analysis, and writing—review and editing.

Both authors contributed to the development of the manuscript and have read and agreed to the published version.

**Conflicts of Interest**

The authors declare no conflicts of interest.

**Data Availability Statement**

No new datasets were generated or analysed during this study. The study is based on previously published literature and publicly available information cited in the manuscript. The sources supporting the discussion and comparative analyses are identified in the reference list.

Preprint Disclosure

An earlier version of this work was posted on arXiv in 2024 under the title "Genetic Bottleneck and the Emergence of High Intelligence by Scaling-out and High Throughput" (arXiv:2407.08743) [40]. The present manuscript has been substantially revised and rewritten, with changes to its title, structure, literature review, theoretical discussion, methodology description, comparative tables, and treatment of limitations and future research.